# Cryptanalysisof two mutual authentication protocols for low-cost RFID

Mohammad Hassan Habibi[1], Mahmoud Gardeshi[2], Mahdi Alaghband[3]

[1]Faculty of Electrical Engineering, I.H. University, Tehran, Iran
mohamad.h.habibi@gmail.com
[2]Faculty of Electrical Engineering, I.H. University, Tehran, Iran
mgardeshi2000@yahoo.com
[3]EEDepartment, Science and Research Campus, Islamic Azad University, Tehran, Iran
m.alaghband@srbiau.ac.ir

## ABSTRACT

*Radio Frequency Identification (RFID) is appearing as a favorite technology for automated identification, which can be widely applied to many applications such as e-passport, supply chain management and ticketing. However, researchers have found many security and privacy problems along RFID technology. In recent years, many researchers are interested in RFID authentication protocols and their security flaws. In this paper, we analyze two of the newest RFID authentication protocols whichproposed by Fu et al. and Li et al. from several security viewpoints. We present different attacks such as desynchronization attack and privacy analysis over these protocols*

## KEYWORDS

*RFID; desynchronization;privacy analysis; mutual authentication; security*

## 1.INTRODUCTION

Radio Frequency Identification (RFID) technology is one of the most important technologies in this decade. This technology allows identifying the tagging objectives wirelessly using transponders queried by readers through a wireless channel. RFID technology has widely been used in applications such as public transportation [1], supply chain management [2], e-passports [3], location tracking systems [4] and access control systems[5].

There are three main components in a RFID system: tags, readers and a backend server. Each tag contains a microchip, antenna and a certain amount of computational and storage capabilities. A reader queries tags to obtain tag contents through wireless communications and sends this information to the backend server through a secure channel. The backend server is composed of a database and some processors [6]. Since the passive tags have low-cost and low computational capabilities, there are information leakage and many security flaws in passive RFID systems. Inasmuch as the passive tags cannot perform complicated cryptography algorithms. The main threats of a RFID system are as following.

- *Tag and reader impersonation*: A malicious adversary masquerades as a legitimate tag and tries to use system services by means of reader deception. On the other hand, a legitimate reader is masqueraded by the attacker and he eventually gets access to the stored secrets of tag [7].
- *Man-in-the middle attack*: As tags and readers use the wireless channel to communicate each other, so this kind of attack can be occurred. In this situation the attacker intervenes between a legal tag and a legitimate reader and exchanges or modifies the authentication messages [8].

- *Tag tracing and tracking*: An adversary traces and tracks legitimate tags from their protocol interactions. The notions *untraceability*, *backward untraceability* and *forward untraceability* are related to this attack [9, 10].
- *Desynchronization*: This is an active attack in which a malicious adversary tries to cause the tag and the reader to update inconsistent values and make tag disabled [11].

In recent years, many researchers have tried to propose lightweight and secure authentication protocols [12, 13, 14, 15, 16, 17, 18,19, 20, 21, 22, 23, 24], but unfortunately many vulnerabilities have been found in their schemes [25, 26, 27, 28, 29, 30, 31, 32, 33, 34, 35, 36, 37, 38]. Recently Fu et al. [39] proposed a scalable RFID mutual authentication and Li et al. [40] suggested a mutual authentication protocol for RFID communication. In this paper, we analyze these protocols and will present three different attacks on FWCFP protocol including desynchronization attack, attack on *untraceability* in two methods and attack on *backward untraceability*. Furthermore, one attack is applied on LWJX protocol which is attack on *untraceability*. The remainder of this paper is organized as following. Related works are studied in section 2.We explain the privacy model for RFID systems in section 3. The FWCFP protocol is summarized as section 4. Our attacks on FWCFP protocol are discussed in section 5. We explain the LWJX protocol in section 6. The security analysis of the LWJX protocol is in section 7 and finally section 8 is assigned to conclusion.The notations in table1 are used throughout this paper.

TABLE 1.THE NOTATIONS

| | |
|---|---|
| $A$ | malicious adversary |
| $E_{K_s}(.)$ | A symmetric encryption function |
| H | a hash function |
| G | a hash function |
| ID | tag identifier |
| IDT | static ID with 96 bit length |
| IDTA | an alias with 96 bit length |
| K | secret value shared by the reader and the tag |
| $K_s$ | Secret key only known by the reader |
| $ID_{old}$ | ID which is used in current communication by the reader after successful authentication |
| $ID_{new}$ | ID which will be used in the new communication by the reader after successful authentication |
| $K_{new}$ | K which will be used in the new communication by the reader after successful authentication |
| $K_{old}$ | K which is used in current communication by the reader after successful authentication |
| $\mathcal{T}$ | the legitimate tag |
| $R$ | The legitimate reader and backend server |
| Rr | random numbers generated by the reader |
| Rt | random numbers generated by the tag |
| $randi(i=0, 1, 2)$ | a random number |
| $X_i^j$ | the item X related to the tag $T_i$ at time $t_j$ |

## 2.RELATED WORKS

In this section we briefly study some authentication protocols which have been proposed to provide secure communications in RFID systems.

Dimitriou proposed an RFID authentication scheme that uses a challenge-response mechanism [39].Since the tag identifier remains constant between two successful sessions, this protocol is vulnerable to tracking attacks and tag impersonation attack.

In [40], a lightweight authentication protocol is proposed by Ohkubo et. Al. This scheme provides indistinguishability and forward security characteristics. The scheme is based on a hash chain and uses two dissimilar hash functions $H$ and $G$. This protocol does not provide protection against an adversary that tries to de-synchronize the server and the tags, consequently resulting in a DoS attack.

Juels [36] showed that cloning and counterfeiting attacks are applied simply on EPC tags. He proposed an unclonable authentication protocol to solve these problems. However, Duc et al. [20] have presented some weaknesses related to privacy and information leakage in Juels scheme.

In [41], Karthikeyan and Nesterenko suggested a security protocol without complex cryptographic primitives. Only XOR and matrix operations were used in their scheme. Chien and Chen [12] showed that this protocol is vulnerable to replay attacks and does not assure the *untraceability* property.

A mutual authentication protocol under the EPC C-1 G-2 standard was proposed by Chien and Chen [12]. They had used simple XOR, CRC and PRNG in their scheme. In [12] each tag needs to keep an EPC code and two secret keys $K_i$, $P_i$. Secret key $K_i$ is used to tag authentication and secret key $P_i$ is used to reader authentication. Both $K_i$ and $P_i$ are updated in each round whereas *EPC* code is permanent. For each tag secret values $K_{old}$, $P_{old}$, $K_{new}$, $P_{new}$, *EPC* and *DATA* are stored in database. The protocol is initialed with sending a random number $N_R$ by the reader. As a result, the tag replies with ($M1$, $N_T$) where $M1=CRC(EPC\|N_R\|N_T)\oplus K_i$. After receiving the tag's response, the database searches for finding the correct tag and its corresponding information ($\{K_{old}, P_{old}\}$ or $\{K_{new}, P_{new}\}$). Thenthe database computes $M2=CRC(EPC\|N_T)\oplus P_x$ ($x= old$ or $new$) and sends tag $M2$. At that point the database updates its secret keys as following: $K_{old}=K_{new}$, $P_{old}=P_{new}$, $K_{new}=PRNG(K_{new})$ and $P_{new}=PRNG(P_{new})$. The tag receives $M2$ and checks whether $M2\oplus P_i=CRC(EPC\|N_T)$. If it satisfies, the tag authenticates the database and updates $K_i$ and $P_i$ the same as with the database, else it terminates the protocol.

Lopez et al. [37] showed some weaknesses of Chien and Chen's protocol including tag and reader impersonation and desynchronization attack. They also showed that this protocol does not guarantee forward security and it is vulnerable to tracing attack. Han and Kwon [14] also presented a desynchronization attack and two tag impersonation attacks on Chien and Chen's protocol in new methods. These attacks were mainly based on weak secure properties of CRC.

## 3.RFIDUNTRACEABLE PRIVACY MODEL

Some privacy models have been proposed by researchers to evaluation of RFID protocols [9, 42, 43, 44]. In [42], Juels and Weis gave a formal definition of the privacy and untraceability model. The same definition is described by Ouafi and Phan in their work presented in ISPEC'08 [44] and we will use this model to analyze the SRP protocol. The model that has been described in [44] is summarized as follows.

The protocol parties are tags ($\mathcal{T}$) and readers ($\mathcal{R}$) which interact in protocol sessions. In this model an adversary $\mathcal{A}$ controls the communication channel between all parties by interacting either passively or actively with them. The adversary $\mathcal{A}$ is allowed to run the following queries:

**Execute** ($\mathcal{R}$, $\mathcal{T}$, $i$) query. This query models the passive attacks. The adversary $\mathcal{A}$ eavesdrops on the communication channel between $\mathcal{T}$ and $\mathcal{R}$ and gets read access to the exchanged messages between the parties in session $i$ of a truthful protocol execution.

**Send** ($\mathcal{U}$, $\mathcal{V}$, $m$, $i$) query. This query models active attacks by allowing the adversary $\mathcal{A}$ to impersonate some reader $\mathcal{U} \in \mathcal{R}$ respectively tag $\mathcal{V} \in \mathcal{T}$ in some protocol session $i$ and send a message $m$ of its choice to an instance of some tag $\mathcal{V} \in \mathcal{T}$ respectively reader $\mathcal{U} \in \mathcal{R}$). Furthermore the adversary $\mathcal{A}$ is allowed to block or alert the message $m$ that is sent from $\mathcal{U}$ to $\mathcal{V}$ (respectively $\mathcal{V}$ to $\mathcal{U}$) in session $i$ of a truthful protocol execution.

**Corrupt** ($\mathcal{T}$, $K'$) query. This query allows the adversary $\mathcal{A}$ to learn the stored secret K of the tag $\mathcal{T} \in \mathcal{T}$, and which further sets the stored secret to $K'$. **Corrupt** query means that the adversary has physical access to the tag, i.e., the adversary can read and tamper with the tag's permanent memory.

- **Test** ($i$, $\mathcal{T}_0$, $\mathcal{T}_1$) query. This query does not correspond to any of $\mathcal{A}$'s abilities, but it is necessary to define the untraceability test. When this query is invoked for session $i$, a random bit $b \in \{0, 1\}$ is generated and then, $\mathcal{A}$ is given $\mathcal{T}_b \in \{\mathcal{T}_0, \mathcal{T}_1\}$. Informally, $\mathcal{A}$ wins if he can guess the bit $b$.

*Untraceable privacy* (*UPriv*) is defined using the game $\mathcal{G}$ played between an adversary $\mathcal{A}$ and a collection of the reader and the tag instances. The game $\mathcal{G}$ is divided into three following phases:

**Learning phase:** $\mathcal{A}$ is given tags $\mathcal{T}_0$ and $\mathcal{T}_1$ randomly and he is able to send any **Execute**, **Send** and **Corrupt** queries of its choice to $\mathcal{T}_0$, $\mathcal{T}_1$ and reader.

**Challenge phase:** $\mathcal{A}$ chooses two fresh tags $\mathcal{T}_0$, $\mathcal{T}_1$ to be tested and sends a **Test** ($i$, $\mathcal{T}_0$, $\mathcal{T}_1$) query. Depending on a randomly chosen bit b $\in \{0, 1\}$, $\mathcal{A}$ is given a tag $\mathcal{T}_b$ from the set $\{\mathcal{T}_0, \mathcal{T}_1\}$. $\mathcal{A}$ continues making any **Execute**, and **Send** queries at will.

**Guess phase:** finally, $\mathcal{A}$ terminates the game $\mathcal{G}$ and outputs a bit b' $\in \{0, 1\}$, which is its guess of the value of b.

The success of $\mathcal{A}$ in winning game $\mathcal{G}$ and thus breaking the notion of UPriv is quantified in terms $\mathcal{A}$ advantage in distinguishing whether $\mathcal{A}$ received $\mathcal{T}_0$ or $\mathcal{T}_1$ and denoted by $\mathbf{Adv}_A^{\mathbf{UPriv}}$ (k) where k is the security parameter.

$\mathbf{Adv}_A^{\mathbf{UPriv}}$ (k) $= \mid \mathrm{pr}\ (b = b') - \mathrm{pr}\ (\text{random flip coin}) \mid = \mid \mathrm{pr}\ (b' = b) - \frac{1}{2} \mid$  where

$0 \leq \mathbf{Adv}_A^{\mathbf{UPriv}}$ (k) $\leq \frac{1}{2}$.

## 4. FWCFP PROTOCOL

Fu et al. proposed a RFID private mutual authentication in [45]. We summarize the proposed protocol as follows. IDT and K are static ID and key with 96 bit length which are shared between each tag and the reader. Each tag also has an IDTA which is an alias with 96 bit length. The reader has a symmetric encryption function $E_{k_s}(.)$ with secret key $K_s$ which is known only by it. The reader uses $E_{k_s}(.)$ to encrypt and decrypt IDTA. In each execution of protocol, IDTA is updated as IDTA = $E_{k_s}$(IDT ‖ rand0) where *rand0* is a random number generated by the reader. The steps of the proposed protocol are as following.

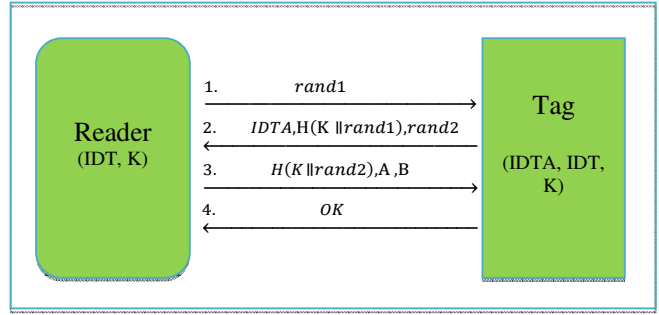

Figure 1. THE FWCFP PROTOCOL

1) The reader generates a random number *rand1*, and sends it to the tag.

2) The reader generates a random number *rand1*, and sends it to the tag.

3) The tag generates a random number *rand2*,computes H(K ‖ *rand1*) and sends {IDTA, H(K ‖ *rand1*),rand2} to the reader. H (.) is a secure hash function.

4) The reader decrypts IDTA using the secret key $k_s$to get the permanent ID of tag-IDT, and then retrieves the shared key K between the tag and the reader by IDT. It computes H (K ‖ *rand1*) and checks whether the computed value equals to the received one. If it matches, the tag is authenticated, otherwise the authentication has failed. If the tag is authenticated successfully, the reader generates a new random number $rand0'$, computes IDTA′ as:

$$\text{IDTA}' = E_{k_s}(\text{IDT} \| rand0') \qquad (1)$$

Then the reader computes the values A and B as:

$$\text{A} = \text{IDTA}' \oplus \text{H(K} \| rand\ 1 \| \text{ra}nd\ 2) \qquad (2)$$

$$\text{B} = \text{IDTA}' \oplus \text{H(K} \| rand2 \| rand1) \qquad (3)$$

It also computes H (K ‖ *rand2*) and sends (H (K ‖ *rand2*), A, B) to the tag.

5) The tag checks H(K ‖ *rand2*) to authenticate the reader. If it matches, the reader is authenticated; otherwise the whole authentication has failed. If the reader is authenticated successfully, the tag computes H(K ‖ *rand 1* ‖ ra*nd 2*) and H(K ‖ *rand 2* ‖ *rand1*). Then it computes two new aliases as:

$$\text{IDTA1}= \text{A} \oplus \text{H(K} \| rand1 \| rand2) \qquad (4)$$

$$\text{IDTA2}= \text{B} \oplus \text{H(K} \| rand2 \| rand1) \qquad (5)$$

If IDTA1= IDTA2, the tag stores IDTA1 as the new alias IDTA and sends OK to the reader.

## 5.SECURITY ANALYSIS OF THE FWCFP PROTOCOL

In this section, we analyze the FWCFP protocol [45] from the security point of view. We have found many security vulnerabilities in this protocol,so we present four different attacks on synchronization and untraceability of this protocol.

### 5.1 Attack on Synchronization

We have found a fundamental weakness in this protocol. An attacker can exploit from this weakness and desynchronize a legal tag $\mathcal{T}_i$ and the legitimate reader. The procedure of the attack is as following.

1) The adversary eavesdrops a valid session between the legal tag $\mathcal{T}_i$ and the reader. He lets parties send the first and the second message safely, but he changes the third message and modifies the values A, B to A′, B′ as:

$$A' = A \oplus \text{IDTA}' \qquad (6)$$

$$B' = B \oplus \text{IDTA}' \qquad (7)$$

where IDTA′ is an arbitrary bit string with 96 bit length. Then the adversary sends (A', B', H(K ∥ rand2)) to Ti as the third message.

2) Upon receiving the third message, $\mathcal{T}_i$ computes H(K ∥ *rand2*), checks whether the computed value equals to the received one. Because it matches, $\mathcal{T}_i$ authenticates the adversary and computes IDTA1, IDTA2 as:

IDTA1 = A′⊕ H (K ∥ *rand 1* ∥ *rand 2*) = A ⊕ IDTA′⊕H (K ∥ *rand 1* ∥ *rand2*) = IDTA′⊕H(K ∥*rand1*∥*rand2*)⊕IDTA′⊕H(K∥*rand1*∥*rand2*)=IDTA⊕IDTA′

$$(8)$$

IDTA2 = B'⊕H(K∥ *rand 2* ∥ *rand 1*) = B ⊕ IDTA′⊕H( K ∥ *rand 2* ∥*rand1*)=IDTA′⊕H(K ∥ *rand2*∥*rand1*)⊕IDTA′⊕H(K∥*rand2*∥*rand1*)=IDTA⊕IDTA′

$$(9)$$

Because IDTA1 = IDTA2, $\mathcal{T}_i$ updates the stored IDTA as:

$$\text{IDTA} = \text{IDTA1} = \text{IDTA}\oplus\text{IDTA}' \qquad (10)$$

At the next sessions, whenever $\mathcal{T}_i$ wants to authenticate itself to the reader, it sends IDTA ⊕IDTA′ to the reader. After decryptionIDTA ⊕ IDTA , the reader extracts IDT' which is not equal to IDT, so the reader does not find IDT' in its database, therefore the reader always rejects $\mathcal{T}_i$ and they have no way to resynchronization.

### 5.2 Attack on Untraceability

A main weakness in designing this protocol is the fact that the term H (shared key ∥ a random number) has the same structure in the second and the third flow of the protocol. An adversary can exploit this weakness and trace a tag as following.

**Learning phase:** The adversary is given tag $\mathcal{T}_0$ at random. A eavesdrops a perfect session between $\mathcal{T}_0$ and a legitimate reader. He gets the values *rand2* and H ($K_0$ ∥ *rand2*) from the first and the second flows of the protocol respectively by an **Execute query**. He reserves these values.

**Challenge phase:** A is given tag $\mathcal{T}_b \in \{ \mathcal{T}_0, \mathcal{T}_1 \}$ randomly. He starts a new session with $\mathcal{T}_b$ and sends *rand2* to it as the first message by **Send query**. $\mathcal{T}_b$ responds with:

(H ($K_b \parallel rand2$), IDTA, $rand'2$) and the adversary reserves H ($K_b \parallel rand2$).

**Guess phase:** If H ($K_b \parallel rand2$) = H ($K_0 \parallel rand2$), the adversary outputs $b' = 0$ and guesses $\mathcal{T}_0'$, otherwise he outputs $b' = 1$ and guesses $\mathcal{T}_1$. The advantage of the adversary is:

$$Adv_A^{Upriv}(k) = \mid pr(A \; wins) - pr(random \; coin \; flip) \mid = \left| pr(b' = b) - \frac{1}{2} \right| = \left| (1 - 2^{-n}) - \frac{1}{2} \right| = \frac{1}{2} - 2^{-n} \quad \text{Where } |H(.)| = n \tag{11}$$

By having H ($K_0 \parallel rand2$), if $\mathcal{T}_b' = \mathcal{T}_0'$, then with the probability of 1 we have H ($K_b \parallel rand2$) = H($K_0 \parallel rand2$), but if $\mathcal{T}_b = \mathcal{T}_1$, then with the probability of $2^{-n}$ we have H ($K_b \parallel rand2$) = H ($K_0 \parallel rand2$), because H (.) is a bit string with length n.

### 5.3 Attack on Backward Untraceability

We use the notion backward untraceability from [8] and use the privacy model from [44] to show that the FWCFP protocol doesn't assure the *backward untraceability*.

**Learning phase:** A is given tag $\mathcal{T}_0'$ at random, he sends $\mathcal{T}_0'$ **Corrupt query** at time $t_1$ and gets the secrets of the $\mathcal{T}_0'$ at time $t_1$ as ( $K_0^1, IDT_0^1, IDTA_0^1$).

**Challenge phase:** The adversary is given $\mathcal{T}_b' \in \{\mathcal{T}_0', \mathcal{T}_1'\}$ randomly. He can have access to the previous session accomplished between $\mathcal{T}_b'$ and R at time $t_0 < t_1$. He gets *rand1* and H ($K_b \parallel rand1$) by **Execute query**.

**Guess phase**: Because the secret key of $\mathcal{T}_0'$ is fixed, we have $K_0^1 = K_0^0$. The adversary also has *rand1* from the session accomplished at time $t_0$. So he can compute H($K_0 \parallel rand1$). Now, if H($K_b \parallel rand1$) = H($K_0 \parallel rand1$), he outputs $b' = 0$ and guesses $\mathcal{T}_0'$, otherwise he outputs $b' = 1$ and guesses $\mathcal{T}_1$. The advantage of the adversary is:

$$Adv_A^{Upriv}(k) = \mid pr(A \; wins) - pr(random \; coin \; flip) \mid = \left| pr(b' = b) - \frac{1}{2} \right| = \left| (1 - 2^{-n}) - \frac{1}{2} \right| = \frac{1}{2} - 2^{-n} \quad \text{Where} |H(.)| = n \tag{12}$$

Because the adversary can compute H($K_0 \parallel rand1$), he owns this value. By having H ($K_0 \parallel rand1$), if $\mathcal{T}_b' = \mathcal{T}_0'$, then with the probability of 1 we have H ($K_b \parallel rand1$) = H ($K_0 \parallel rand1$), but if $\mathcal{T}_b = \mathcal{T}_1$, then with the probability of $2^{-n}$ we have H ($K_b \parallel rand1$) = H ($K_0 \parallel rand1$), because H (.) is a bit string with length n.

## 6. LWJX PROTOCOL

Li et al. proposed an authentication protocol for secure RFID communication [46]. The proposed protocol is as it follows.Each tag stores an initial ID and a secret key K shared by the tag and the reader. The reader keeps the following information for each tag: ID with initial value same as to tag's ID, hash value of $ID_{new}$ with the initial value of H (ID), hash value of$ID_{old}$ with the initial value is empty, new value of secret key $K_{new}$ with the initial value of K, old value of secret key $K_{old}$ with the initial value is empty. The parameter M holds howmany times a tag has had unsuccessful sessions. Two hash functions H and G are implemented on each tag and on the reader. The procedure of authentication is as follows.Each tag stores an initial ID and a secret key K shared by the tag and the reader. The reader keeps the following information for each tag: ID with initial value same as to tag's ID, hash value of $ID_{new}$ with the initial value of H (ID), hash value of$ID_{old}$ with the initial value is empty, new value of secret key $K_{new}$ with the initial value of K, old value of secret key $K_{old}$ with the initial value is empty.The parameter M holds howmany times a tag has had unsuccessful sessions. Two hash functions H and G are implemented on each tag and on the reader. The procedure of authentication is as follows.

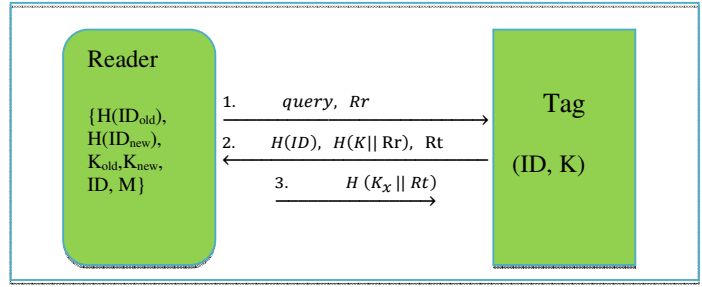

Figure 2. THE LWJX PROTOCOL

1) The reader generates a number Rr at random and sends it to the tag.
2) After receiving Rr, the tag generates a number Rt at random, computes H (ID) and H (K || Rr), then it sends them and Rt to the reader.
3) The reader searches in the database records tofind whether there is a H ($ID_{old}$), H ($ID_{new}$) equal to the received H (ID). Three possible cases occur:
   a) If no value is found, it terminates the protocol.
   b) If it is found that H($ID_{new}$) = H (ID), then the readercomputes H ($K_{new}$ || Rr) and compares it with the received H (K||Rr), if H ($K_{new}$ ||Rr)≠ H (K||Rr), then R terminates the protocol; otherwise,Rresets M=0, computes H ($K_{new}$ || Rt) and sends H ($K_{new}$ || Rt) to the tag. Finally, the reader updates its secret values as following:

$$ID = G (ID) \qquad (13)$$

$$H(ID_{old}) = H(ID_{new}) \qquad (14)$$

$$H(ID_{new}) = H(ID) \qquad (15)$$

$$K_{old} = K_{new} \qquad (16)$$

$$K_{new} = ID \oplus Rr \oplus Rt \qquad (17)$$

   c) If H($ID_{old}$) = H(ID) , first the reader checks M,if M isgreater than the upper limit, it terminates the protocol and issues a warning;  otherwise it makes M= m+1, now if H($K_{old}$ ||Rr)≠ H($K_{old}$||Rr), then R terminates the communication; otherwise,Rcomputes H($K_{old}$|| Rt) and sends H($K_{old}$|| Rt) to the tag. The reader doesn't update in this case.
4) After receiving H ($K_x$||Rt) {x=old or new}, ifH ($K_x$||Rt)=H (K||Rt), the tagupdates its secretvalues as:

$$ID = G(ID) \qquad (18)$$

$$K = ID \oplus Rr \oplus Rt \qquad (19)$$

## 7. PRIVACY ANALYSIS OF LWJX PROTOCOL

In this section, we analyze the LWJX protocol and give an attack on this protocol.

### 7.1 Attack on Untraceability

We give our privacy analysis on LWJX protocol according to the privacy model discussed in [18] which has been explained it in section III. We show that the LWJX protocol does not have *untraceability*.

**Learning phase:** The adversary is given tag $\mathcal{T}_0$ at random. He masquerades as a legitimate reader and starts a new session with tag $\mathcal{T}_0$ by sending **Send query**. He sends $Rr_1$ to $\mathcal{T}_0$ and gets its response as (H $(ID_0)$, H $(K_0 \parallel Rr_1)$, Rt). The adversary reserves these values and terminates the session to avoid the $\mathcal{T}_0$ updating.

**Challenge phase:** A is given $\mathcal{T}_b \in \{\mathcal{T}_0, \mathcal{T}_1\}$ randomly. The adversary performs a new session with $\mathcal{T}_b$ by sending **Send query**. He sends $Rr_1$ to $\mathcal{T}_b$ and gets its response as (H$(ID_b)$, H($K_b \parallel Rr_1$), $Rt'$). The adversary reserves these values and terminates the session.

**Guess phase:** The adversary can guess the correct tag in two ways:

If H($ID_b$)= H($ID_0$) , then the adversary outputs b'=0 and guesses $\mathcal{T}_0$; otherwise he outputs b'=1 and guesses $\mathcal{T}_1$.

1) If H($K_b \parallel Rr_1$) = H ($K_0 \parallel Rr_1$), then the adversary outputs b'=0 and guesses $\mathcal{T}_0$, otherwise he outputs b'=1 and guesses $\mathcal{T}_1$. In both cases, A wins with high probability:

$$\boldsymbol{Adv}_A^{Upriv}(k) = \mid pr \, (A \, wins) - pr \, (random \, coin \, flip)\mid \, = \left| pr \, (b' = b) - \frac{1}{2} \right| =$$

$$\left|(1 - 2^{-n}) - \frac{1}{2}\right| = \frac{1}{2} - 2^{-n} \text{ Where } \mid H \, (.)\mid = n \qquad (20)$$

Because |H(.)| = n, we have H($ID_0$)=H($ID_1$) with the probability of $2^{-n}$, so the adversary can guess the correct tag with the probability of $1 - 2^{-n}$.

### 8. CONCLUSION

In this paper, we showed some security and privacy vulnerabilities of the RFID authentication protocols proposed by Fu et al [45] and Li et al [46]. We also present the desynchronization attack and tag tracing on [45]. In desynchronization attack, an adversary can easily change the third message transmitted in protocol and desynchronize the target tag and the legitimate reader. We also presented the privacy analysis of this protocol in a formal privacy model. We showed the FWCFP protocol doesn't assure *untraceability*, *backward untraceability* and *forward untraceability*. We also presented some attacks on privacy and anonymity of [46]. It has shown that *untraceability* and *forward untraceability* aren't assured by this protocol.

### ACKNOWLEDGMENT

This work is supported by the Education & Research Institute for ICT, Tehran, Iran.

# REFERENES

**Authors**

Mohammad Hassan Habibi received the Bachelor's degree in Telecommunication Engineering From Kerman University, Kerman, Iran, in 2007 and Master's degree in Telecommunication in the field of Cryptography with the honor degree from IHU, Tehran, Iran (2011), where he obtained the Best Student Academic Award.
Currently, he is a research assistant (RA) at the research center of cryptography, IHU, Tehran, Iran. His research interest includes: Lightweight cryptography, RFID security, authentication protocols,cryptanalysis, public key cryptography and lightweight primitives.

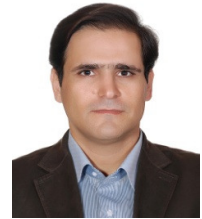

Mahmoud Gardeshi received his Erudition Degree in applied mathematics from Amir KabirUniversity, Islamic Republic of Iran in 2000. Currently, he is a researcher at the I. H. University. His research interest includes: cryptography and information security.

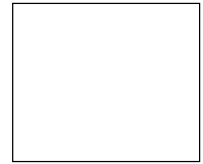

Mahdi R. Alaghband received his B.S. degree in Electrical engineering in 2005 and M.S. degree in Communications, Cryptology & Information Security in 2008. Currently, he is both a Ph.D. candidate at the department of Electrical and Computer Engineering, Azad University and research assistant on Information Systems and Security Lab (ISSL), EE Dept., Sharif University of Technology. His research interests include authentication protocol especially in RFID systems, security in wireless sensor networks and lightweight cryptographic.

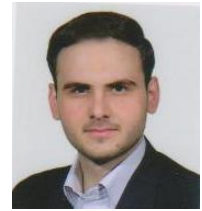